# Visual Path Tracking Control for Park Scene

Linjiong Zhu, Wenfu Wang, Weijie Yang, Zhijie Pan, An Chen
College of Computer Science and Technology, Zhejiang University

## ABSTRACT

Autonomous driving application is developing towards specific scenes. Park scene has features such as low speed, fixed routes, short connection, less complex traffic, and hence is suitable for bringing autonomous driving technology into reality. This paper targets park scene, and proposes a visual path tracking lateral control method using only one webcam. First, we calculate error of distance and error of angle from camera images, and then use fuzzy logic to fuzzify them into a combined error degree. The PID control algorithm takes it as input, and outputs steering wheel angle control command. Fuzzification could tolerate the error brought by image transformation and lane detection, making PID control more stably. Our experiments in both virtual and real scene show that our method can accurately and robustly follow the path, even at night. Compared with pure pursuit, our method can make 5 meters turning.

## 1. INTRODUCTION

Autonomous driving is promised to have evolutionary impact on transportation system, even the whole society. Many countries and companies are developing this technology. Despite millions of miles of autonomous driving road tests by Google, Apple and so on, L5 level autonomous driving is actually far beyond reach. In *The Hype Cycle* published by Gartner in 2017, L5 autonomous driving is in the *Peak of Inflated Expectations* phase, and it will take more than ten years for this technology to mature. Just lately, Uber self-driving vehicle killed a pedestrian while road testing, this arouses worldwide discussion about the safety of autonomous driving technology. Many companies like Toyota and Nvidia suspended their self-driving road tests right after this accident.

Autonomous driving is facing many challenging difficulties in complicated traffic scene. For one thing, a self-driving car is usually equipped with very expensive sensors, such as LiDAR, high-precision GPS, and so on. These expensive sensors are expected to make self-driving cars safe and sound, while Uber and Tesla's accidents tell us another story. Second, cameras and LiDAR generate huge amount of data, it requires workstations to process these data in real time. Moreover, the complex traffic scenarios are endless. Millions of miles road test cannot guarantee its safety. All these make autonomous driving very difficult to move on to people's everyday life.

In fact, autonomous driving application is developing towards specific scenes. In specific scenes, many features are favorable for self-driving technology to happen. These features are low speed, short connection, fixed routes, less complicated traffic and so on. In fact, many companies are going this way. For example, a company called Navya is running autonomous shuttles on fixed routes in campus and hospital. SB drive provide service in Japan rural area for elders. Similar companies like EasyMile and UISEE are also focusing on specific scenes. These vehicles use less expensive LiDAR and computers, but is still much more expensive than regular cars.

Autonomous driving application requires safe and robust technology with reasonable cost. More importantly, we believe the self-driving application scenario should be redesigned for it to work. In a develop zone park, we redesign the short connection self-driving scenario, and manage to run trial operation. In our redesign, lane markers of routes are painted on the road, besides two lanes markers, we also add another lane marker in the center of road. People are not allowed to enter the road. The route is fixed, with a number of stops for passengers to get onto the self-driving car. Passengers order cars through an APP online, and are expected to arrive reserved stops on time. Then passengers scan the QR code to start their journey.

This paper focuses on the visual path tracking control method applied to our short connection self-driving application, and we will discuss our redesigned self-driving application in another paper. It has merits such as accurately and robust path following, path following at night, small radius turning, all with reasonable cost.

## 2. BACKGROUND

Path tracking control is well-studied in robotics. Path tracking refers to a vehicle executing a pre-defined geometric path by applying appropriate steering motions that guide the vehicle along that path. According to this definition, a good path tracking controller will minimize error of distance and error of angle. Given different localization sensors, we divide path tracking controller into two categories: localization-based path tracking and vision-based path tracking.

### 2.1 Localization-based Path Tracking

In robotics, a robot needs to know its precise location in its surrounding environment to choose its action. Usually, high-precision GPS or LiDAR could provide absolute global position. Given the path, the controller output actions to follow the path.

One of the most popular path tracking controllers is pure pursuit[1]. Pure pursuit is a geometric path tracking controller, and it builds on vehicle kinematic bicycle model. It exploits geometric relationship between the robot's pose and the given path. Specifically, pure pursuit finds the furthest point from itself on the given path within a look-ahead distance, and then taking rear wheel as center and velocity direction as tangent. It calculates the curvature which takes itself back to the given path. Finally, it outputs the steering wheel angle given the curvature and vehicle kinematic bicycle model. Pure pursuit is simple and effective, however, look-ahead distance is difficult to set. Large look-ahead distance results in cutting corners, small brings about instability of tracking performance. Many researches focus on improving pure pursuit. Ollero[2] adopted fuzzy logic to tune the look-ahead distance. The Stanley method[3], which helped Stanford won the DARPA Urban Challenge, computed the heading error and the shift error between the center of front axle and the nearest path point. And it then transferred the shift error to angle metric with a tangent function. The output steering angle is addition of the two errors.

Control theory controllers are also based on precise localization, Zhao[4] used adaptive PID controller for path tracking. While PID controller suffers from parameters setting and overshot in tracking performance. Snider[5] employed Linear Quadratic Regulator(LQR) to track the path. Falcone[6] used Model Predictive Control(MPC) approach, and experimented under extreme conditions. MPC could perform robustly on icy roads with 21 m/s speed, which proved its powerful performance. MPC and LQR could promise fine accuracy, but need more computational resources to optimize.

In summary, localization-based path tracking controllers requires precise position to accurately execute path following. High-precision GPS or LiDAR could provide accurate position information, however these sensors are very expensive. Besides, high-precision GPS often lose some of the satellites signal in cities, resulting large positioning error. On the other hand, most of localization-based controller builds on vehicle kinematic bicycle model, which assumes small steering angle, therefore these controllers' performance deteriorates when making small radius turn which is common in park scene.

### 2.2 Vison-based Path Tracking

As embedded devices are gaining more computing power and cameras are cheaper, more and more researches utilize vision as input for path tracking control[7][8]. Vision-based path tracking task is defined as guiding the motion of a robot with respect to a target path based on the feedback obtained through a vision system. According to building map or not, we can divide them into map-based and mapless path tracking approaches.

In map-based method, a map is first built and stored using visual SLAM or other methods to extract visual features. When testing, images captured by cameras are matching with pre-built map to localize itself. Roye [9] first recorded videos along some path, and the vehicle is driven by people. Then, they built the path and environment in 3D map, after that the vehicle can drive along the same path autonomously. The problem with map-based method is that visual map suffers from poor precision, which cannot provide precise position, the tracking accuracy is therefore poor. Another problem is that feature matching process is slow and consuming a lot of resources, which makes this method unstable.

Mapless path tracking methods do not require prior knowledge about the environment or pre-built map[10]. It acquires visual features as control error variable from vision system to execute feedback control to guide the vehicle along the path. We believe this approach is most suitable for path tracking task in our redesigned park scene. Previous research attempts are briefly described below. Zhi[11] first detected lane markers from images, and then used PID[12] to control vehicle. They calculate distance error and direction error at preview distance in the image, however they require high-precision error values. Environment factors such as daylight and whether severely affect visual detecting, this leads to various errors when calculating distance and angle errors. In [13], fuzzy control was used to tune the PID parameters. Li[14] used visual preview and fuzzy sliding mode control for path tracking.

Our method takes advantages of previous research methods which combines fuzzy logic and visual detecting. Fuzzy logic tolerates errors induced by calculating distance and angle error, and makes PID control more stable. We will detail our method in the next section.

## 3. VISUAL PATH TRACKING

In this paper, we propose a visual path tracking control method based on monocular camera. The framework of our approach is shown in Figure 1, the details are in the subsections.

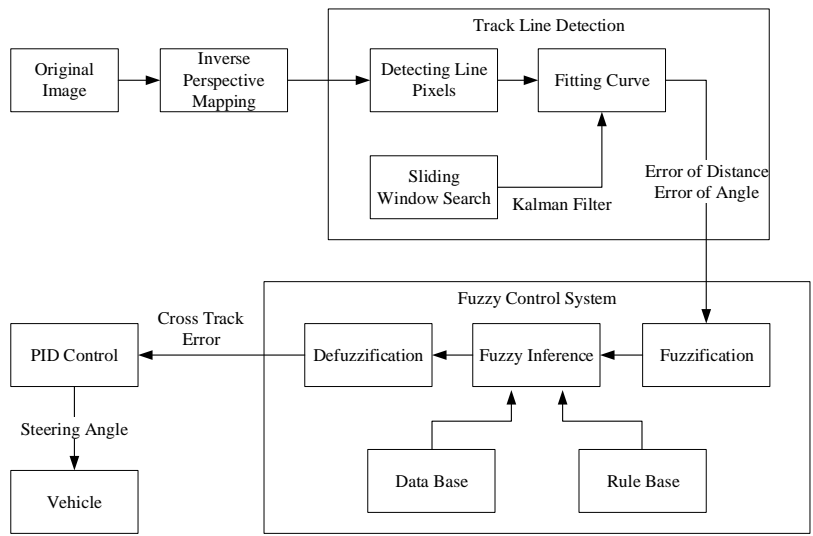

Figure 1 Framework of the approach

### 3.1 Inverse Perspective Mapping

First, the perspective of the original image from the camera is wrapped to an overhead view of the road. As shown in Figure 2, the world coordinate system W and the original image coordinate system I are defined to describe the transformation of inverse perspective mapping[15].

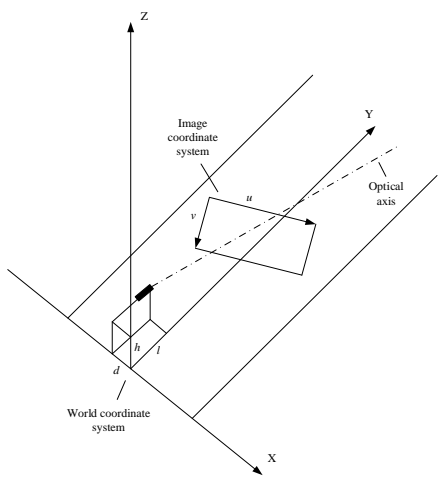
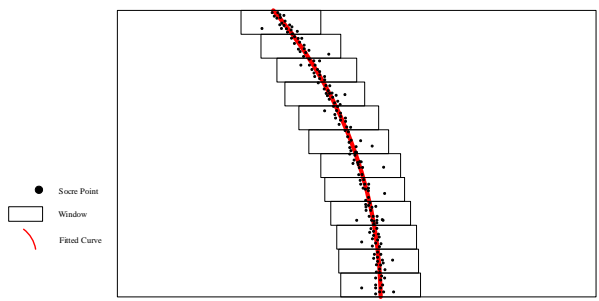

Figure 2 Inverse perspective mapping     Figure 3 Sliding window search for fitting curve

Assuming that, the position coordinates of the camera is (*d, l, h*) in *W*. The parameter, $\gamma$, is the angle between the projection line of the camera's optical axis at the $z=0$ plane and y-axis. Another parameter, $\theta$, is the deviation angle of the camera's optical axis to the $x=0$ plane. The field of view angle of the camera in the horizontal directions is $2\alpha_u$, and the field of view angle of the camera in the vertical directions is $2\alpha_v$. In the image of camera, $R_u$ is horizontal resolution and $R_v$ is vertical resolution.

The inverse perspective transform model representing the coordinate transformation from *I* to *W* is shown in equation (3-1) and (3-2). But the transformation are nonlinear and not bidirectional one-to-one mapping, and there are blank pixels in the resultant image[16]. Thus, a reverse IPM transform model, which is shown in equation (3-3) and (3-4), is adopted to stuff the resultant image.

$$x(u,v) = h \times \cot\left(\frac{2\alpha_v}{R_v - 1} \times v - \alpha_v + \theta\right) \times \sin\left(\frac{2\alpha_u}{R_u - 1} \times u - \alpha_u + \gamma\right) + d \tag{3-1}$$

$$y(u,v) = h \times \cot\left(\frac{2\alpha_v}{R_v - 1} \times v - \alpha_v + \theta\right) \times \cos\left(\frac{2\alpha_u}{R_u - 1} \times u - \alpha_u + \gamma\right) + l \tag{3-2}$$

$$v = \frac{(R_v - 1) \times \left(\arctan\frac{h}{\sqrt{(x-d)^2 + (y-l)^2}} + \alpha_v - \theta\right)}{2\alpha_v} \tag{3-3}$$

$$u = \frac{(R_u - 1) \times \left(\arctan\frac{(x-d)}{(y-l)} + \alpha_u - \gamma\right)}{2\alpha_u} \tag{3-4}$$

### 3.2 Track Line Detection

#### 3.2.1 Detecting Line Pixels

Whether a pixel is selected depends on its value in appointed channels of color spaces, which are LAB B, HSV value, and HLS lightness color channels.

The steps are as follows. 1) Translate the overhead view image into LAB, HSV, HLS color space, and select the appointed channels, which are LAB B, HSV value, and HLS lightness color channels. 2) In each selected channel, normalize overhead view image by CLAHE (Contrast Limited Adaptive Histogram Equalization) [17], and strong local contrast and correction for shadows are ensured. 3) Create a binary image by selecting for pixels only above a certain intensity. 4) Combine the result from different channels into a result image, and the intensity is highest where the color channels agree on line pixel locations.

#### 3.2.2 Fitting Curve

As shown in Figure 3, the result image is divided into 12 parts along the height direction and a window with width of W is set in every bisection. In each bisection, Kalman filter method is used to update the location of the window, based on the distribution of line pixels in this bisection and the location of the window last time. The Kalman filter method not only avoids the influence of noise points, but also predicts the location of window when the track line is temporarily disappearing, so as to ensure the continuity of the window in time and the continuity of the action. After updating all windows, a polynomial fit is applied to create the curve.

We calculate the error of distance (EOD) and error of angle (EOA) from the curve. EOD is defined as the difference between the abscissa of the far point of the curve and the abscissa of the image center point. EOA is defined as the tangential angle of the far point of the curve.

### 3.3 Fuzzification

Fuzzy control system use fuzzy logic that analyzes analog input values in terms of logical variables of continuous values between 0 and 1 [18][19]. In this paper, calculation of values of EOD and EOA are not accurate. In order to reduce the effect of visual measurement's error, EOD and EOA are put into a fuzzy system, and output a new value, called cross track error (CTE). CTE is the input of the following PID control system.

Assuming that, the fuzzy set of EOD and EOA are both {NB, NM, NS, ZO, PS, PM, PB}, whose universes are both **R**. The fuzzy set of CTE is {NBX, NB, NMB, NM, NMS, NS, ZO, PS, PMS, PM, PMB, PB, PBX}, whose universe is **R**. This paper defines membership functions of EOD, EOA and CTE are shown in Figure 4, considering of the code for design of urban road engineering [20]. The rules of fuzzy inference are defined as Table 1.

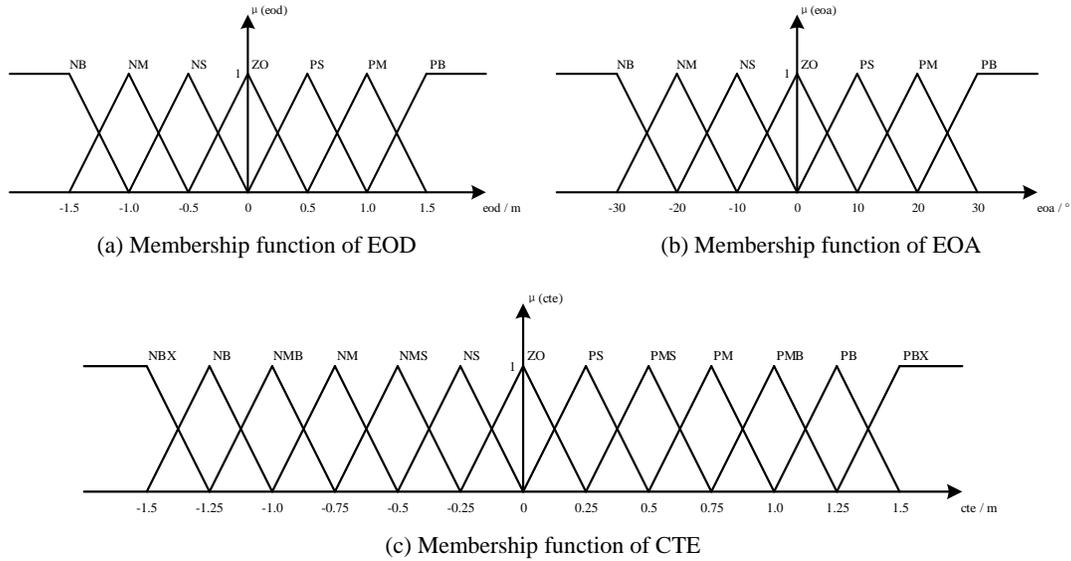

(a) Membership function of EOD

(b) Membership function of EOA

(c) Membership function of CTE

Figure 4 Definition of membership functions

Table 1 Rules for fuzzy inference

| EOA / EOD | NB | NM | NS | ZO | PS | PM | PB |
|---|---|---|---|---|---|---|---|
| NB | NBX | NB | NMB | NM | NMS | NS | ZO |
| NM | NB | NMB | NM | NMS | NS | ZO | PS |
| NS | NMB | NM | NMS | NS | ZO | PS | PMS |
| ZO | NM | NMS | NS | ZO | PS | PMS | PM |
| PS | NMS | NS | ZO | PS | PMS | PM | PMB |
| PM | NS | ZO | PS | PMS | PM | PMB | PB |
| PB | ZO | PS | PMS | PM | PMB | PB | PBX |

The steps in Fuzzy control system is as follows. 1) Fuzzify EOD and EOA according to Figure 4, 2) Change the fuzzified values of EOD and EOA into the fuzzified value of CTE by rules in Table 1, 3) Defuzzify the value of CTE by center-of-gravity method, whose mathematical model is shown as the equation (3-5). In the equation, $U$ is the accurate CTE, and $u_i$ is the $i$ th fuzzified CTE, whose degree of membership is $K_i$.

$$U = \frac{\sum_{i=1}^{n} u_i K_i}{\sum_{i=1}^{n} u_i} \quad (i=1, 2, ..., 13) \tag{3-5}$$

## 3.4 PID Control

In this paper, an increment-based PID is used to output steering wheel angle to control the vehicle. The input of PID control system is CTE, which is the output of fuzzy system. In this way, the PID system controls more stably, which is validated in experiments of section 4.

Steps in PID control system are as followed.

(1) Before starting the PID control system, three parameters are set, which are $K_p$, $K_i$, and $K_d$.

(2) Assuming that, the $n$ th CTE is $cte[n]$. If $n$ is equal to 1, $cte[n-1]$ and $cte[n-2]$ are set as 0.

(3) Update $p_{error}$, $i_{error}$, $d_{error}$, $cte[n-2]$, $cte[n-1]$ as equation (3-6) ~ (3-10).

$$p_{error} = cte[n] - cte[n-1] \tag{3-6}$$
$$i_{error} = cte[n] \tag{3-7}$$
$$d_{error} = cte[n] - 2 \times cte[n-1] + cte[n-2] \tag{3-8}$$
$$cte[n-2] = cte[n-1] \tag{3-9}$$
$$cte[n-1] = cte[n] \tag{3-10}$$

(4) Calculate $steering\_angle$ as equation (3-11), and send it to the vehicle.

$$steering\_angle = -(K_p \times p_{error} + K_i \times i_{error} + K_d \times d_{error}) \tag{3-11}$$

## 4. EXPERIMENTS

We first develop in a virtual park scene created with Unity3D, after experiments in virtual scene, we then implement in a development zone park scene. The virtual scene guides our design of real world scene. The virtual scene and development zone park scene are shown in Figure 5. We experiment different road curvatures and vehicle speed in virtual scene, and find a suitable road curvature of 8 m. The development zone park scene has 8 m road curvature, and total distance of about 550 m.

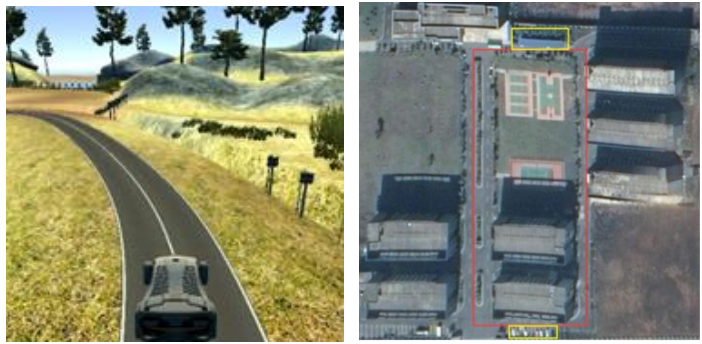

Figure 5 Virtual Scene and Development Zone Park Scene

### 4.1 Virtual Scene Experiments

The virtual scene has two lanes and another central lane. The camera is a front facing camera at the height of 0.8m, whose field of view is 60 degrees. The width and height of the image from camera are 320x240 pixels. In virtual scene experiment, we verify the function of fuzzy system, and test a 5 m radius turning.

#### 4.1.1 Fuzzy Control System

In order to validate the importance of fuzzy control system for stability tracking performance, we set an experiment, called "simulation experiment with fuzzy control" (SE-FC), and another called "simulation experiment with no fuzzy control" (SE-NFC). In the two experiments, vehicle speed is both 20km/h, and PID parameters are both set as $K_p=10.0$, $K_i=10.0$, and $K_d=4.0$.

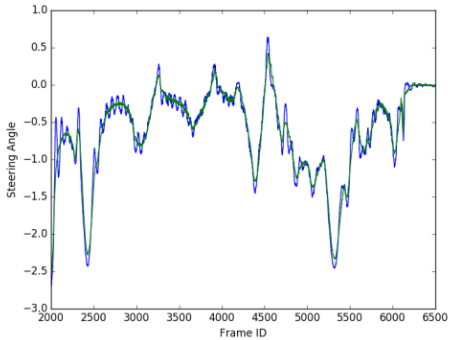 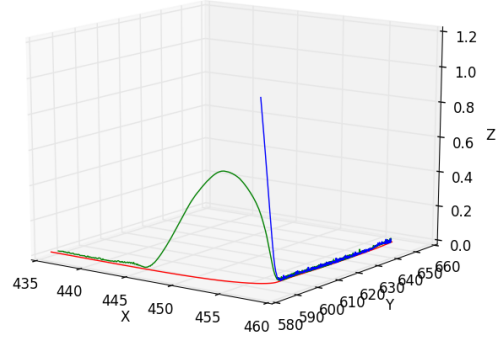

Figure 6 Steering angle results of SE-FC and SE-NFC       Figure 7 Track result of SET-V and SET-PP

The steering angle results of SE-FC and SE-NFC are shown as Figure 6 and Table 2. The abscissa is the frame ID, and the ordinate is steering angle whose unit is degree. The green one shows the result of SE-FC and the blue one shows the result of SE-NFC. Obviously, fuzzy control system makes the fluctuation of steering angle smaller, and make the PID control system more stable. This results a more smooth and accurate path tracking process. Besides, the average process time for one frame is 43.61ms in SE-FC and 39.79ms in SE-NFC.

Table 2 Experiment results

| Experiment Name | Scene Type | Minimal Turning Radius (m) | Maximal CTE (m) | Average Process Time for One Frame (ms) |
|---|---|---|---|---|
| SE-FC | Virtual Scene | 20 | 2.372 | 43.61 |
| SE-NFC | Virtual Scene | 20 | 2.518 | 39.79 |
| SET-V | Virtual Scene | 5 | 0.610 | 42.86 |
| SET-PP | Virtual Scene | 5 | >1.022 | 50.00* |
| E-V | Park Scene | 8 | 1.727 | 72.19 |
| E-VN | Park Scene | 8 | 1.733 | 72.21 |

*Remarks: In SET-PP, this paper set location frequency as 20Hz for pure pursuit.

### 4.1.2 Small Radius Turning

We test a 5 m radius turn in virtual scene, the experiment is called "simulation experiment for turning with vision" (SET-V) using our visual path tracking controller. As a baseline, we also conduct pure pursuit method to track the same path, this experiment is called "simulation experiment for turning with pure pursuit" (SET-PP). In the two experiments, vehicle speeds are both 10km/h, and PID parameters for our method is $K_p=10.0$, $K_i=14.0$, and $K_d=1.0$, and the look-ahead distance fort pure pursuit is 8 m.

The results of SET-V and SET-PP are shown as Figure 7 and Table 2. The XY plane is the path plane, and Z axis shows the cross track error whose unit is meter. The red one shows the road. The green one shows the result of SET-V and the blue one shows the result of SET-PP. When turning, the maximum cross track error of SET-V is about 0.5m. As for SET-PP, the cross track error is increasing and the vehicle gets out of the lane. This experiment shows that our approach has the ability to make small radius turns. As for average process time, the result is 42.86ms in SET-V and 50ms in SET-PP.

### 4.2 Park Scene Experiments

After experiments in virtual, we paint the road in park scene which has a center lane and turning radius of 8 m. We successfully finish our path using our method on the vehicle platform. Our vehicle platform is shown is Figure 8, it is an electric car with steer-by-wire and brake-by-wire control. The process computer is Nvidia Jetson TX2 and a camera mounted at the height of 0.8m, and with the field of 60 degrees view. The image we get from this camera is 320x240 pixels, which is the same as in the virtual scene.

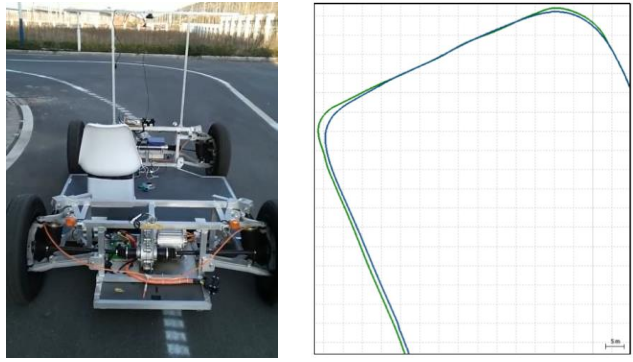

Figure 8 Park scene experiments

Using our method, our vehicle accurately follow the path in the development zone park scene with 10 km/h. The PID parameters are set as $K_p = 20.0$, $K_i = 18.0$, and $K_d = 1.0$. The results are shown in Figure 8 and Table 2. The experiment is called "experiment with vision" (E-V). The green trace is the result of E-V and the blue trace is path we want to track. We can see that our method performs very accurate path following.

In addition, we also make experiment at night called "experiment with vision at night" (E-VN). It is shown in Figure 9 and Table 2. In E-VN, the LED of vehicle is turned on, and other settings are the same as E-V. The vehicle can maintain its path following ability as in daytime. In E-V and E-VN, the average process time is similar, which is about 72ms.

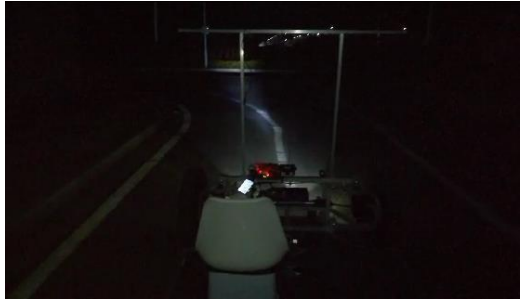

Figure 9 Park scene experiments at night

## 5. CONCLUSION

In this paper, an accurate and stable visual path tracking control approach for park scene is proposed. The advantages of our approach are as follows. 1) A fuzzy control system is used to tolerate the errors induced by image calculation, and has an improvement of accuracy and stability for PID control system. 2) Our method has the ability of small radius turning, which is favorable for park scene. 3) Our method outperforms pure pursuit method in both virtual and real scene, and achieve it with much lower price. We conduct experiments in both virtual and park scene to verify our method.

## ACKNOWLEDGE

The work is supported by National Natural Science Foundation of China under Grants (61751209).

# REFERENCES

[1] Coulter R C. Implementation of the pure pursuit path tracking algorithm[R]. Carnegie-Mellon UNIV Pittsburgh PA Robotics INST, 1992.

[2] Ollero A, García-Cerezo A, Martínez J L. Fuzzy supervisory path tracking of mobile reports[J]. Control Engineering Practice, 1994, 2(2): 313-319.

[3] Thrun S, Montemerlo M, Dahlkamp H, et al. Stanley: The robot that won the DARPA Grand Challenge[J]. Journal of field Robotics, 2006, 23(9): 661-692.

[4] Zhao P, Chen J, Song Y, et al. Design of a control system for an autonomous vehicle based on adaptive-pid[J]. International Journal of Advanced Robotic Systems, 2012, 9(2): 44.

[5] Snider J M. Automatic steering methods for autonomous automobile path tracking[J]. Robotics Institute, Pittsburgh, PA, Tech. Rep. CMU-RITR-09-08, 2009.

[6] Falcone P, Borrelli F, Asgari J, et al. Predictive active steering control for autonomous vehicle systems[J]. IEEE Transactions on control systems technology, 2007, 15(3): 566-580.

[7] Cherubini A, Chaumette F, Oriolo G. An image-based visual servoing scheme for following paths with nonholonomic mobile robots[C]//Control, Automation, Robotics and Vision, 2008. ICARCV 2008. 10th International Conference on. IEEE, 2008: 108-113.

[8] Bonin-Font F, Ortiz A, Oliver G. Visual navigation for mobile robots: A survey[J]. Journal of intelligent and robotic systems, 2008, 53(3): 263-296.

[9] Royer E, Bom J, Dhome M, et al. Outdoor autonomous navigation using monocular vision[C]//Intelligent Robots and Systems, 2005.(IROS 2005). 2005 IEEE/RSJ International Conference on. IEEE, 2005: 1253-1258.

[10] Güzel M S. Autonomous vehicle navigation using vision and mapless strategies: a survey[J]. Advances in Mechanical Engineering, 2013, 5: 234747.

[11] Zhi-qiang H. Design of smart car section PID control algorithm based on CCD camera[J]. Electronic Design Engineering, 2011, 2: 024.

[12] Ang K H, Chong G, Li Y. PID control system analysis, design, and technology[J]. IEEE Transactions on Control Systems Technology, 2005, 13(4):559-576.

[13] Thuy P X, Cuong N T. Vision Based Autonomous Path/Line Following of a Mobile Robot Using a Hybrid Fuzzy PID Controller[J]. PROCEEDING of Publishing House for Science and Technology, 2016, 1(1).

[14] Li L, Lian J, Wang M, et al. Fuzzy sliding mode lateral control of intelligent vehicle based on vision[J]. Advances in Mechanical Engineering, 2013, 5: 216862.

[15] Deng J, Han Y. A real-time system of lane detection and tracking based on optimized RANSAC B-spline fitting[C]// Research in Adaptive and Convergent Systems. 2013:157-164.

[16] Wang C, Shi Z K. A Novel Traffic Stream Detection Method Based on Inverse Perspective Mapping[J]. Procedia Engineering, 2012, 29:1938-1943.

[17] Pizer S M, Amburn E P, Austin J D, et al. Adaptive histogram equalization and its variations[J]. Computer Vision Graphics & Image Processing, 1987, 39(3):355-368.

[18] Hájek P. Metamathematics of Fuzzy Logic[J]. Trends in Logic, 1998, 4:155-174.

[19] Pedrycz W. Fuzzy Control and Fuzzy Systems[M]// Fuzzy control and fuzzy systems. Research Studies Press, 1989:97-98.

[20] CJJ 37-2012, Code for design of urban road engineering [S].